\documentstyle[12pt,epsf]{article}

\title{Comment on "Constraining a possible dependence of Newton's constant on
the Earth's magnetic field"}

\author{J.  P.  Mbelek \\ Service d'Astrophysique, C.E.  Saclay \\ F-91191
Gif-sur-Yvette Cedex, France}

\begin{document} \maketitle \baselineskip=8mm

\begin{abstract} Recently A.  Rathke has argued that the KK$\psi$ model
explanation of the discrepant measurements of Newton's constant is already ruled
out due to E\"otv\"os experiments by several orders of magnitude.  The structure
of the action of the KK$\psi$ model is even qualified as inconsistent in the
sense that it would yield a negative energy of the electromagnetic field.  Here,
I refute both claims and emphasize the possibility still open to reconcile the
experimental bounds on the test of the weak equivalence principle (WEP) with
scalar-tensor theories in general by some compensating mechanism.
\end{abstract}

\section{Introduction} In a recent paper~\cite{Rathke}, A.  Rathke has expressed
some criticisms on the KK$\psi$ model~\cite{mbeleka} - \cite{mbelekc} by
performing a rough estimate of the metric perturbation of a single nucleus and
the effective contribution to the gravitational mass it may involve.  The author
claimed that such analysis can be applied to generic theories with gravielectric
coupling.  Thus, he concluded that the bounds of WEP violations put by the
E\"otv\"os experiments rule out the explanation of the discrepant measurements
of Newton's constant by such couplings by several orders of magnitude.
Furthermore, Rathke argued that the very structure of the action of the KK$\psi$
model is inconsistent in the sense that it would yield a negative energy of the
electromagnetic field.  Consequently, according to him, the action of the
KK$\psi$ model is just a mixture of contributions written in different
conventions leading to the misinterpretation that the Kaluza-Klein (KK) theory
is classically unstable.  However, both claims, as stated yet, are incorrect in
many respects.  In the next section we recall the major arguments we have put
forward to conclude the instability of the genuine KK theory.  In Sec.  3, we
discuss the validity of the procedure used by Rathke to make a conclusion on WEP
violation by the KK$\psi$ model.  In Sec.  4, we present a new opportunity to
prevent KK-like theories or scalar-tensor theories from any significant WEP
violation.  A proof is given in the appendix.

\section{Instability of the genuine 5D KK theory} Rathke argues that our claim
of the instability of the genuine 5D KK theory at the classical level is a
misconception due to the conventions we have employed.  Further, he suggests to
us to refer to Ref.~\cite{Wehus} for a didactic derivation and discussion of the
KK action in various frames.  In this respect, I reiterate that we have shown
the instability of the genuine KK theory independently of the frame
(see~\cite{mbelekd}).  In particular, this is obvious in Einstein frame (one
passes from the Jordan frame to Einstein frame by the conformal transformation
${\hat{g}}_{AB} \rightarrow {\Phi}^{-1/3} \,{\hat{g}}_{AB}$, $A, B = 0, 1, 2, 3,
4$ are 5D labels, whereas $\alpha, \beta = 0, 1, 2, 3$ are 4D labels), as can be
seen in the literature~\cite{Overduin, Iyer} - \cite{Appelquist} and displayed
below :  \begin{equation} \label{KK action Pauli} S_{KK} = - \,\int \sqrt{- g}
\,(\frac{R}{{\kappa}^{2}} \, \,+ \,\frac{1}{4} \, \Phi \,F^{\alpha\beta}
\,F_{\alpha\beta} \,+ \,\frac{1}{6{\kappa}^{2}{\Phi}^{2}} \,{\partial}^{\alpha}
\Phi \,{\partial}_{\alpha} \Phi) \,d^{4}x \end{equation} to be compared to
\begin{equation} \label{minimal scalar} S = - \,\int \sqrt{- g} \,(
\,\frac{R}{{\kappa}^{2}} \, \,+ \,\frac{1}{4} \,F^{\alpha\beta}
\,F_{\alpha\beta} \,- \,\frac{1}{2} \,{\partial}^{\alpha} \phi
\,{\partial}_{\alpha} \phi \,+ \,U(\phi) \,+ \,J \,\phi) \,d^{4}x,
\end{equation} in the case of a classical real scalar field, $\phi$, minimally
coupled to gravity with the potential $U = U(\phi)$ and the source term $J =
J(x^{\alpha})$, where we have set ${\kappa}^{2} = 16\pi G/c^{4}$.  As regards
the featuring of a kinetic term for the 5D KK scalar field, $\Phi$, which is
usually absent in the Jordan frame, the reader may refer, {\sl e.  g.},
to~\cite{Overduin} (Sec.  3.3, Eq.(9)) for opposing views.  In any case, the
absence of a kinetic term just makes the 5D KK theory with zero electromagnetic
field equivalent to a Brans-Dicke theory with parameter $\omega = 0$.  Now, as
shown by Noerdlinger~\cite{Noerdlinger}, stability of the Brans-Dicke action
requires $\omega > 0$.  The limiting case $\omega = 0$ leads to an unstable
vacuum \cite{Constantinidis}.  Thus, our conclusion on the instability of the
genuine KK theory still holds both in the Einstein frame (whose action meets a
common consensus) and in the Jordan frame even if the KK scalar field kinetic
term is removed from the action.

\section{A proof of WEP violation ?}

First, the definition of the (effective) gravitational mass given
in~\cite{Rathke} is wrong and quite confusing.  The author is manifestly dealing
with the gravitational self-energy of a nucleus and not the gravitational mass
itself, this inconsistency appearing clearly in the lack of a second mass
density and a double integral in Eqs.(41), (42) of~\cite{Rathke} in contrast
with Eq.(2) of ref.~\cite{Nordtvedta} (Sec.  2).  Rathke does not seem to
realise that the WEP violation occurs in scalar-tensor theories only because of
the spatial variation of the binding energies of composit particles.  Moreover,
in the spirit of E\"otv\"os experiments, generally one considers a WEP violation
in the external gravitational field of a macroscopic body by comparing
accelerations of two test bodies of various composition ({\sl e.  g.}, the Earth
and the Moon in the gravitational field of the Sun~\cite{Nordtvedta}.  A reason
for that is the weakness of gravity as compared to the strength of other
fundamental interactions like the electromagnetic interaction.  In fact, it
seems that Rathke just tried to evaluate the effective gravitational self-force
of a nucleus and its influence on the WEP violation.  However, it is not so
simple to compute the gravitational self-force since one should deal with gauge
and regularization problems which are by far out of reach of a simple
phenomenological approach (see~\cite{Nakano}).  Furthermore, referring to the
recent approach known as the "chameleon" cosmology, scalar fields like those of
the KK$\psi$ model can acquire a huge mass in region of high mean density, as it
is indeed the case within a nucleus in contrast with the atmospheric density or
vacuum in usual laboratory experiments~\cite{Khoury} (see also~\cite{Nojiri} for
stability considerations).

\section{A possibility of compensating WEP violation}

Like any physical model, the KK$\psi$ model may finally turn out not to be
viable.  However, it is worth noticing that the problem raised by the discrepant
G measurements still remains.  Moreover, the correlation that we have
established from experimental data between the geomagnetic field and the
laboratory measurements of G at different locations on Earth is still
unchallenged.  On the other hand, we agree with Rathke that E\"otv\"os-like
experiment put very strong constraints on any violation of the universality of
free fall (UFF).  Nevertheless, this does not necessarily imply that large
gravielectric couplings are inconsistent with known experimental bounds on the
WEP, as we show below.  Actually, the claim that KK-like theories and
scalar-tensor theories in general should violate the WEP is not new at all.  It
goes back to Jordan and Dicke themselves~\cite{Dicke}.  Nevertheless, there
still remains a possibility of a mechanism that could help such models or
variants of such models to conciliate a large gravielectric coupling with the
WEP.  Thus, in the framework of his version of varying fine structure constant
(which too couples the Maxwell invariant $F^{\alpha\beta} \,F_{\alpha\beta}$ to
a scalar field), Bekenstein has suggested a cancellation mechanism in order to
escape any WEP violation due to the different internal constitutions of
objects~\cite{Bekenstein}.  However, his proposal was soon after strongly
criticized by Damour~\cite{Damour}.  Let us recall that there was a time when
the quark model was thought to be ruled out by the Pauli principle (which is at
least as fundamental as the WEP) because of the existence of the resonance
$\Delta^{++}$.  In that case, a symmetry, the color, was a solution.  Keeping
this in mind, let us explore in what follows a new possibility which may
conciliate the experimental bounds on WEP violation with the KK-like models
(see~\cite{mbelekb}).  Following Nordtvedt~\cite{Nordtvedtb}, a body whose mass
depends on some parameter $\alpha$ which varies in space will be subject to a
body-dependent acceleration \begin{equation} \label{extra acceleration}
\vec{\delta a} = \,- \,\frac{\partial \ln{M}}{\partial \ln{\alpha}} \,\,c^{2}
\,\,\vec{\nabla} \ln{\alpha} \end{equation} which accounts for UFF violation.
Hence, two different bodies, labelled $i$ and $j$, of different composition will
fall in the Earth's gravitational field $\vec{g}$ with a difference in
accelerations whose magnitude equals \begin{equation} \label{eq motion 5}
\frac{\mid \vec{a}_{i} \,- \,\vec{a}_{j} \mid}{\mid \vec{g} \mid} =
\frac{\partial \ln{(M_{i}/M_{j})}}{\partial \ln{\alpha}} \,\,c^{2}
\,\,\frac{\mid \vec{\nabla} \ln{\alpha} \mid}{g} \end{equation}

Consequently, although the mass $M = M(\alpha)$ of a given body will depend on
the variable parameter $\alpha = \alpha(\Phi(\vec{r}), \psi(\vec{r}))$, there
will be no observed UFF violation for $M = M_{0} \,K(\Phi, \psi)$, where $M_{0}$
is a positive constant and $K(\Phi, \psi)$ is a universal function which reduces
to unity when the scalar fields are not excited.  We show in what follows that
the Higgs mechanism of dynamical generation of elementary particles'mass may
also allow the latter possibility.  Indeed, let us consider the mass $M$ of a
given atom on account of its composition, namely, $Z$ protons and $N$ neutrons
in the nucleus, plus $Z$ surrounding electrons.  It is written in the general
form :  \begin{equation} \label{atom mass formula} M = ( \,2Z \,+ \,N \,)
\,m_{u} \,+ \,( \,2N \,+ \,Z \,) \,m_{d} \,+ \,Z\,m_{e} \,- \,\frac{\Delta
E}{c^{2}} \end{equation} where \begin{equation} \label{binding energy} \Delta E
= {\alpha}_{eff}^{2} \,F_{em}(Z, N) \,+ \,{\alpha}_{s~eff}^{n_{s}} \,F_{s}(Z, N)
\,+ \,{\alpha}_{w~eff}^{n_{w}} \,F_{w}(Z, N) \end{equation} is the internal
binding energy which includes the electromagnetic interaction, the strong
interaction and the weak interaction binding energies contributions,
respectively.  The subscripts $em$, $s$ and $w$ stand for the electromagnetic,
strong and weak interactions, respectively; the quantities ${\alpha}_{eff},
{\alpha}_{s~eff}, {\alpha}_{w~eff}$ are the corresponding relevant effective
coupling constants.  The exponents $n_{s}$ and $n_{w}$ need not to be specified
for our purpose.  Now, the masses $m_{u}$, $m_{d}$ and $m_{e}$ result
dynamically from the Yukawa coupling of the $u-$quark, the $d-$quark and the
electron to the Higgs field.  Now, in our framework, the Yukawa coupling
constants $\{g^{(u)}, g^{(d)}, g^{(e)}\}$ of these elementary particles should
be replaced with the effective quantities $\{g^{(u)}_{eff}, g^{(d)}_{eff},
g^{(e)}_{eff}\}$ which all depend on both scalar fields $\Phi$ and $\psi$ (see
Appendix).  Clearly, the only way to get rid of a composition-dependent ratio
$M/M_{0}$ consists in relating the effective coupling constants to each other so
that \begin{equation} \label{coupling constants connexion}
\frac{g^{(u)}_{eff}}{g^{(u)}} = \frac{g^{(d)}_{eff}}{g^{(d)}} =
\frac{g^{(e)}_{eff}}{g^{(e)}} = \left( \frac{{\alpha}_{w~eff}}{{\alpha}_{w}}
\,\right)^{n_{w}} = \left( \frac{{\alpha}_{s~eff}}{{\alpha}_{s}}
\,\right)^{n_{s}} = \left( \frac{{\alpha}_{eff}}{\alpha} \,\right)^{2} = K(\Phi,
\psi), \end{equation} where the sets of quantities $\{g^{(u)}, g^{(d)},
g^{(e)}\}$ and $\{{\alpha}_{w}, {\alpha}_{s}, \alpha\}$ denote the Yukawa
coupling constants and the fundamental interaction coupling constants,
respectively, when both scalar fields are not excited.  A connection between the
ratios ${\alpha}_{w~eff}/{\alpha}_{w}$, ${\alpha}_{s~eff}/{\alpha}_{s}$ and
${\alpha}_{eff}/\alpha$ has already been put foreward in~\cite{mbelekc} (see the
discussion), on the basis of the fundamental interactions unification scheme.
Usually, the phenomenological formula given in the textbooks reads
\begin{equation} \label{empirical formula} \Delta E = a_{v}\,A \,-
\,a_{s}\,A^{2/3} \,- \,a_{c}\,\frac{Z^{2}}{A^{1/3}} \,- \,a_{a}\,\frac{( \,N \,-
\,Z \,)^{2}}{4A} \,- \,\delta\,A^{1/2}, \end{equation} where $A = Z \,+ \,N$ and
the values of the parameters $a_{v}$, $a_{s}$, $a_{c}$, $a_{a}$ and $\delta$
depend on the range of masses for which they are optimized (see, {\sl e.
g.},~\cite{Povh et al.}).  Hence, according to our analysis, the parameters
$a_{v}$, $a_{s}$, $a_{c}$, $a_{a}$, $\delta$ should be considered as effective
parameters that scale as $K(\Phi, \psi)$.  This is still consistent with the
natural expectation that these parameters should depend on the fundamendal
interactions coupling constants.

\section{Conclusion} The Kaluza-Klein theory is unstable even in the Jordan
frame, irrespective of whether or not the kinetic term of the KK scalar field is
included in the action functional.  A.  Rathke's claims that the KK$\psi$ model
is already ruled out because of its gravielectric coupling is not yet founded,
based solely on the naive model he has sketched.  The definition of the
gravitational energy which he has used is also incorrect.  Hence, his
conclusions are not based on serious proofs.  Instead, it seems still possible
to accommodate a gravielectric coupling with the UFF on account of the Higgs
mechanism of dynamical mass generation.

\section{Appendix} The Lagrangian density of a fermion (quark or lepton) of mass
$m$ whose wave function is the spinor $\Psi$ reads \begin{equation}
\label{spinor Lagrangian} L = \frac{i}{2} \,[ \,\bar{\Psi}
\,{\gamma}^{\alpha}\,D_{\alpha}\,\Psi \,- \,( \,D_{\alpha}\,\bar{\Psi} \,)
\,{\gamma}^{\alpha} \,\Psi \,] \,- \,m c \,\bar{\Psi} \,\Psi, \end{equation}
where $D_{\alpha}$ stands for the covariant derivative and hereafter $\hbar =
1$.  Noether's theorem yields the energy-momentum density (see, {\sl e.
g.},~\cite{Leite Lopes}, Sec.  I.  7, p.  49 and Sec.  I.  8, p.  60)
\begin{equation} \label{spinor energy-momentum} T^{\alpha\beta} = \frac{i}{2}
\,[ \,\bar{\Psi} \,{\gamma}^{\alpha}\,D^{\beta}\,\Psi \,- \,(
\,D^{\beta}\,\bar{\Psi} \,) \,{\gamma}^{\alpha} \,\Psi \,] \,+ \,m c
\,\bar{\Psi} \,\Psi \,g^{\alpha\beta}, \end{equation} on account of the Dirac
equation.  Then, it follows \begin{equation} \label{spinor energy-momentum
trace} T^{\alpha}_{\alpha} = 4m c \,\bar{\Psi} \,\Psi.  \end{equation} In the
KK$\psi$ models (see~\cite{mbelekc}, Sec.  $2.2$), fermions will generate the
$\psi$-field through a source term of the form \begin{equation} \label{source
term from Psi} J = \frac{8\pi G}{3 c^{4}} \,g(\Phi, \psi) \,T^{\alpha}_{\alpha}
= \frac{32\pi G}{3 c^{4}} \,g(\Phi, \psi) \,m c \,\bar{\Psi} \,\Psi .
\end{equation} The total action reads \begin{equation} \label{total action} S =
\,S_{KK,4} \,+ \,\,\frac{c^{4}}{4\pi G} \,\int \sqrt{-g} \,\,\Phi \,\left[
\,\frac{1}{2} \,{\partial}_{\alpha} \psi \,\,{\partial}^{\alpha} \psi \,\,-
\,\,U \,\,- \,\,J \psi \,\right] \,d^{4}x \,+ \,\int \sqrt{-g} \,L \,d^{4}x,
\end{equation} where $S_{KK,4}$ denotes the genuine 5D KK action after
dimensional reduction.  As can be seen, the relation (\ref{source term from
Psi}) provides the spinor $\Psi$ with an effective mass $m_{eff} = m \,K(\Phi,
\psi)$ which is expressed as \begin{equation} \label{effective mass of Psi}
m_{eff} = m \,\left[ \,1 \,+ \,\frac{8}{3} \,g(\Phi, \psi) \,\psi \,\Phi
\,\right].  \end{equation} As one knows, the mass of an elementary fermion is
derived from the Yukawa coupling of $\Psi$ to the Higgs field, $\phi$, by
replacing the mechanical mass $m$ in Eq.  (\ref{spinor Lagrangian}) with
$g^{(\Psi)} \,\phi$ (see~\cite{Leite Lopes}, Sec.  VIII.  3 and \cite{Rolnick,
Bilenky}), where $g^{(\Psi)}$ is the Yukawa coupling constant of the fermion.
Thus, Eq.  (\ref{effective mass of Psi}) involves an effective Yukawa coupling
constant given by \begin{equation} \label{effective Yukawa coupling}
g^{(\Psi)}_{eff} = g^{(\Psi)} \,\left[ \,1 \,+ \,\frac{8}{3} \,g(\Phi, \psi)
\,\psi \,\Phi \,\right].  \end{equation}

\end{document}